# Reconstruction of Sub-Nyquist Random Sampling for Sparse and Multi-Band Signals

Amir Zandieh, Alireza Zareian, Masoumeh Azghani, Farokh Marvasti
Advanced Communication Research Institute (ACRI)
Department of Electrical Engineering
Sharif University of Technology
Tehran, Iran
{zareian,zandie_amir}@ee.sharif.edu, elmaazghani@yahoo.com,
fmarvasti@sharif.edu

*Abstract*— As technology grows, higher frequency signals are required to be processed in various applications. In order to digitize such signals, conventional analog to digital convertors are facing implementation challenges due to the higher sampling rates. Hence, lower sampling rates (i.e., sub-Nyquist) are considered to be cost efficient. A well-known approach is to consider sparse signals that have fewer non-zero frequency components compared to the highest frequency component. For the prior knowledge of the sparse positions, well-established methods already exist. However, there are applications where such information is not available. For such cases, a number of approaches have recently been proposed. In this paper, we propose several random sampling recovery algorithms which do not require any anti-aliasing filter. Moreover, we offer certain conditions under which these recovery techniques converge to the signal. Finally, we also confirm the performance of the above methods through extensive simulations.

## I. Introduction

The uniform sampling theorem states that for low pass signals, we need a sampling rate that is at least twice the highest frequency component of the signal. There are various applications where this rate is costly to achieve, due to the increase of the bandwidth and the complexity of the implementation.

Landau [1] showed that the necessary sampling rate for the reconstruction of the multiband signals is at least twice the total length of the occupied bandwidth. The target of sub-Nyquist sampling is to reconstruct a signal with a sampling rate as low as the Landau rate. Moreover, it has been shown in [2]-[3] that for discrete sparse random signals, we need at least $O(k \log(n/k))$ samples per frame, where $k$ is the sparsity number and $n$ is the signal frame length. There are cases where we know the position of the occupied bands in the frequency domain, i.e., the spectral support. The spectral support information significantly helps in the reconstruction of the signal from its sub-Nyquist samples. Several efforts have been done to achieve the Landau rate using the spectral support information [4]-[26]. However, if we do not know the spectral support, the Landau rate is a challenging bound, and the recovery methods usually need to sample at a multiple of the Landau rate [12]-[25]-[27]. Note that for sparse signals, a multiple of the landau rate can still be much less than the Nyquist rate.

As an application of random sampling, we consider Analog to Digital Converters (ADCs) for multi-band signals. In a radio communication system, to process the received signal, a demodulation technique is necessary to be used before the ADC. This demodulation becomes hard and expensive to implement where multiple carriers are needed to be scanned. This becomes harder for the case of unknown carriers (e.g., military surveillance, radar, and medical imaging applications), and even harder for time-varying carriers (e.g., frequency hopping). Since digital technology is much simpler than the analog equivalent, we would like to digitize the RF spectrum prior to the demodulation stage. However, the Nyquist rate required for digitizing the RF signals is too high, which increases the complexity of the ADC; to address this issue we need to design a sub-Nyquist ADC.

The sub-Nyquist sampling has become very popular in the last decade; Compressed Sensing (CS) [2]-[3], [12]-[19]-[23] is one approach, and random sampling [5]-[11] is another one. The CS method exploits the sparsity property for recovery from a set of linear measurements. A signal is called sparse if most of its coefficients are zero in some domain, such as Discrete Cosine Transform (DCT), Discrete Wavelet Transform (DWT) or Discrete Fourier Transform (DFT) [2]-[3]. If a signal has more than 50 percent nonzero coefficients, typically the signal is called dense. For continuous signals such as RF signals, sparsity is defined when the total occupied bandwidth is much lower than the total band (from the lowest frequency component to the highest frequency component). Sparse signal processing has found enormous applications in a broad range of research fields such as spectrum sensing [11]-[12], sparse channel estimation [13], direction of arrival estimation [14], detection of radar signals [15], and face recognition [16].

Various CS recovery algorithms have been proposed in the literature. The very first techniques were based on L1 minimization such as Basis Pursuit (BP) [17], and Least Absolute Shrinkage and Selection Operator (LASSO) [18], which achieve high precision recovery at the cost of remarkable computational complexity. The greedy algorithms were then suggested to speed up the recovery procedure at the expense of accuracy. Orthogonal Matching Pursuit (OMP) [19] and COSAMP [20] are the well-known examples of this group. The iterative thresholding techniques such as IST [21] and IHT [22] apply a simple recursive relation to present a fast estimation of the signal vector with acceptable accuracy.

A number of ADCs have been designed based on CS to work at sub-Nyquist rate. In [23], an ADC system is offered and implemented on hardware [23]-[24]. In this system, a random demodulation technique which demodulates the signal with a high-rate pseudo-noise is applied on the signal followed by an anti-aliasing filter and a sub-Nyquist uniform sampler. The recovery of the sampled signal is feasible by applying CS recovery algorithms.

Another CS based recovery technique is proposed in [25] which exploits the periodic nonuniform sampling using $m$ independent uniform samplers with random time delays. Modulated Wideband Converter (MWC) [26], [27] combines the ideas of the random demodulation and multicoset sampling techniques. This method modulates the signal in $M$ parallel channels with different pseudo-noises. It uses an anti-aliasing filter and samples the output of each channel uniformly with a sampling rate relatively close to the Landau bound.

The random sampling technique can also exploit the sparsity of a signal at the sub-Nyquist rates. The advantage of random sampling is that we do not need an anti-aliasing filter besides the fact that we are using a sub-Nyquist rate. A hardware design for random sampling is proposed in [28]-[29] using analog multiplexers and an array of switched capacitors or sample-and-hold devices.

In this paper, we aim to address the sub-Nyquist sampling problem based on random sampling of sparse signals. We use a variation of an algorithm developed in our laboratory [11], [30], [31], for the recovery. We also give a formal proof for the convergence of the algorithm. The performance of the proposed technique and its robustness against noise is verified through simulations of synthetic and real signals.

The rest of this paper is organized as follows: Section II discusses the useful characteristics of random sampling and provides essential mathematical formulation. Section III is the main section where we propose our reconstruction and its convergence proof. Section IV is related to the algorithms and parameter settings. Section V includes numerical evaluations and simulation results. Finally, Section VI concludes the paper.

## II. THEORY OF RANDOM SAMPLING

In this section, we analyze the characteristics of random sampling and demonstrate how we can use it as a sub-Nyquist sampling for sparse signals. Fig. 1. illustrates the result of uniform and random sampling of a sparse signal (S) in the DFT domain. Fig. 1. (a) represents a typical sparse signal (S) with the highest frequency component of 9KHz and total bandwidth of 2.5KHz. In this case, the Nyquist rate becomes 18 KHz and the Landau rate is 5 KHz; Fig. 1. (b) demonstrates the result of a uniformly upsampled version of S at 25 KHz; similarly, Fig. 1. (c) is the uniformly downsampled version of S at 15 KHz. The signal in Fig. 1. (d) represents the result of random sampling of S at the average rate of 15 KHz. In contrast to Fig. 1. (b), the signal shown in Fig. 1. (c) shows the aliasing effect, i.e., the interference of the high frequency components as a result of sub-Nyquist sampling.

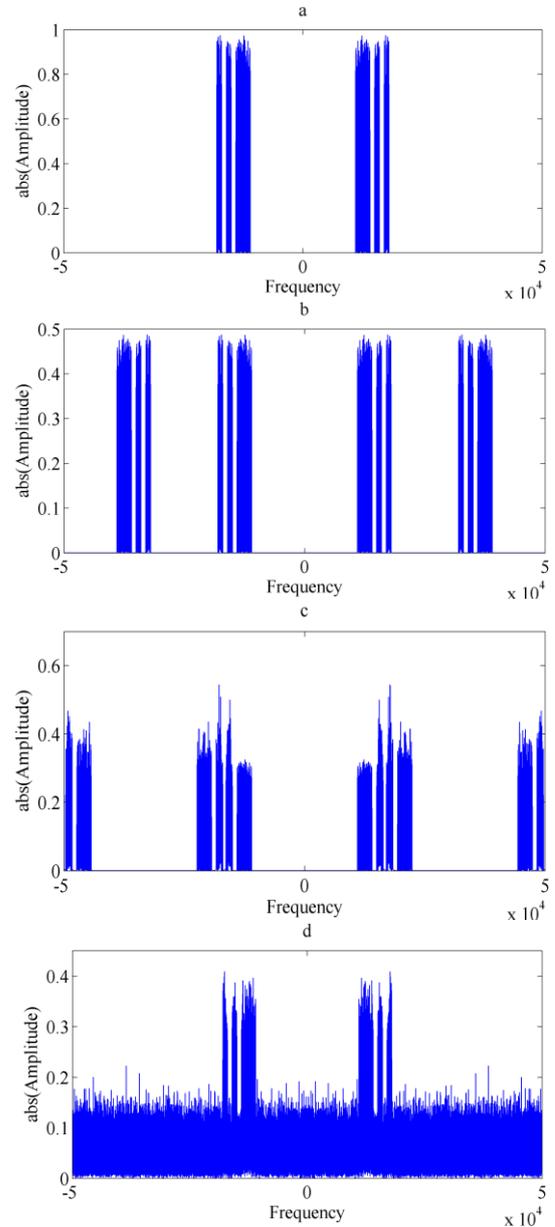

Fig. 1. (a) The main signal, (b) uniformly upsampled, (c) uniformly downsampled (d) randomly sampled at the same downsampled sub-Nyquist rate.

Although Fig. 1. (d) is associated with a sub-Nyquist sampling, all the frequency components are visible and only affected by an additive noise. This figure shows that random sampling preserves the sparsity pattern of the signal buried in a background noise.

To show this formally, suppose that $x(t)$ is the input signal and $x_s(t)$ is the signal randomly sampled with uniform distribution. In other words,

$$x_s(t) = \sum_{i=1}^{m} \delta(t - t_i) x(t) = \psi(t) x(t) \qquad (1)$$

where $t_i$'s are *i.i.d.* random variables with uniform distribution and $\psi(t)$ denotes a random Dirac comb function, i.e., $\psi(t) = \sum_{i=1}^{m} \delta(t - t_i)$, and $m$ denotes the total number of samples. Also note that throughout this paper, subscript $s$ for any function $x_s(t)$ indicates the random samples of $x(t)$.

It is shown in [32]-[33] and [5][5] that for random sampling with uniform distribution and $m \to \infty$, $\psi(t)$ is a stationary stochastic process and its power spectrum is:
$$S_\psi(f) = \lambda^2 \delta(f) + \lambda$$
where $\lambda$ is the sampling rate (the average number of samples in a unit time interval). If $x(t)$ is a stationary stochastic process, $x_s(t)$ will become a stationary stochastic process and:
$$S_{x_s}(f) = S_\psi(f) * S_x(f) = \lambda^2 S_x(f) + \lambda P_x \quad (2)$$

where $S_{x_s}(f)$ and $S_x(f)$ denote the power spectra of $x_s(t)$ and $x$, respectively. $P_x$ denotes the total power of the signal $x(t)$ derived from the integral of $S_x(f)$. Since $\lambda P_x$ is a constant number, it represents the spectrum of a zero-mean white noise; this explains the shape shown in Fig. 1. (d). Hence, $x_s(t)$ can be represented as:

$$x_s(t) = \lambda x(t) + n(t) \quad (3)$$

where $n(t)$, referred to as sampling noise, is a white noise generated due to random sampling with the variance of $\lambda P_x$.

Before proving the convergence of the Iterative Method with Adaptive Thresholding (IMAT), we present a number of lemmas and theorems. Assuming uniform distribution for the samples, $t_i$'s, we analyze the Fourier transform of the comb function. It is easy to show that the Fourier transform of the comb function, $\psi(t)$, is a stochastic process in the form of $\Psi(f) = \sum_{i=1}^{m} e^{-j2\pi f t_i}$ characterized in the following lemma.

*Lemma 1:* The Fourier transform of the comb function is a non-stationary stochastic process in the form of:
$$\Psi(f) = \hat{\Psi}(f) + \lambda \delta(f) \quad (4)$$
where $\hat{\Psi}(f)$ is a stationary zero mean white Gaussian process with autocorrelation:
$$R_{\hat{\Psi}}(F) = \lambda \delta(F)$$
and $\lambda$ is the sampling rate.

The proof of this lemma is straightforward.
Now suppose that $x(t)$ is a deterministic signal, the Fourier transform of $x_s(t)$ yields:
$$X_s(f) = \Psi(f) * X(f)$$

The following theorem clarifies the statistical characteristics of $X_s(f)$.

*Theorem 1:* If $x(t)$ is a deterministic signal, then the Fourier transform of its random samples, $X_s(f)$, would be a stochastic process in the form of:

$$X_s(f) = \lambda X(f) + N_X(f) \quad (5)$$

where $\lambda$ is the sampling rate, $X(f)$ is the Fourier transform of $x(t)$ and $N_X(f)$ is a stationary Gaussian process with the power spectrum:
$$S_{N_X}(\theta) = \lambda |x(-\theta)|^2, \quad \forall \theta \in \mathbb{R} \quad (6)$$

*Proof:* According to Lemma 1, we would have:
$$X_s(f) = [\hat{\Psi}(f) + \lambda \delta(f)] * X(f) = \lambda X(f) + N_X(f)$$

where $N_X(f) = \hat{\Psi}(f) * X(f)$. According to Lemma 1, $\hat{\Psi}(f)$ is a stationary Gaussian process; hence, by the fact that Gaussian distribution is preserved under linear transformation, $N_X(f)$ will become Gaussian and it suffices to compute its power spectrum. First, the autocorrelation of $N_X(f)$ is as:
$$\begin{aligned} R_{N_X}(F) &= R_{\hat{\Psi}}(F) * X(F) * \overline{X(-F)} \\ &= \lambda \delta(F) * X(F) * \overline{X(-F)} \\ &= \lambda X(F) * \overline{X(-F)} \end{aligned} \quad (7)$$

The power spectrum of $N_X(f)$ can be obtained as:
$$S_{N_X}(\theta) = \lambda x(-\theta)\overline{x(-\theta)} = \lambda |x(-\theta)|^2$$
∎

According to the previous theorem, $N_X(f)$ at each frequency $f$ is a Gaussian random variable with the variance:

$$\sigma^2 = P_{N_X} = \int_{-\infty}^{\infty} S_{N_X}(\theta) d\theta = \lambda \int_{-\infty}^{\infty} |x(\theta)|^2 d\theta = \lambda P_x \quad (8)$$

According to (5) and (8), $X_s(f)$ consists of the signal $\lambda X(f)$ contaminated by a Gaussian noise, $N_X(f)$, with variance $\sigma^2 = \lambda P_x$, which confirms the results for the uniformly distributed stochastic point process given in (2). Moreover, we know that the magnitude of a zero-mean Gaussian variable with the variance of $\sigma^2$ is less than $\alpha\sigma$ with the probability of $\Phi(\alpha)$, where $\Phi$ is the cumulative normal distribution function [34]. Thus, the signal $\lambda X(f)$ can be extracted from $X_s(f)$ by thresholding it as follows:
$$T(z) = \begin{cases} z & \text{if } |z| > Thr \\ 0 & \text{else} \end{cases} \quad (9)$$

where:
$$Thr \geq \alpha \sqrt{\lambda P_x}$$

$\alpha$ is chosen to be $\Phi^{-1}(p)$, and $p$ is the acceptable probability for each noise component to be removed. For example, $\alpha$ should be 2.58 for 99% confidence.

Another useful property of $N_X(f)$ is its ergodicity which is investigated in the following lemma.

*Lemma 2:* If $X(f)$ is absolutely integrable, then the stochastic process $N(f)$ will be ergodic.

The proof of this lemma follows from the fact that $N(f)$ is a Gaussian process which is ergodic if and only if its autocorrelation is absolutely integrable. The proof is included in the appendix for completeness.

## III. THE PROPOSED RECOVERY METHOD

In this section, we use the foundation provided in Section II and propose an iterative algorithm for sparse reconstruction of randomly sampled signals. According to Fig. 1. (d), random sampling preserves the sparsity pattern of the original signal and only a background white noise is added to the original signal; this suggests a thresholding technique for the recovery. Before detailed explanation, we demonstrate a simple pseudo-code of the Iterative Method of Adaptive Thresholdin (IMAT) as the core of our approach.

The sampling mask in the following algorithms represents a binary vector which has a value of 1 where there are samples and 0 where there are no samples.

---

**Algorithm 1**: Iterative Method with Adaptive Thresholding (IMAT)

- The capital letter for each symbol refers to the Fourier transform of the signal represented by a small letter.
- $S(j)$ for any signal $E$ means the index $j$ of the vector $S$ and by $j$ we mean any integer number between 1 and $L$.

**Input:**
- A random sampling mask $Mask \in R^n$
- A random sampled signal $y \in R^n$

**Output:**
- A recovered estimate $r \in R^n$ of the original signal

**Procedure IMAT(y, x):**
1. $s^0 = y$
2. For $i = 1 : iter\_max$ do
3. $S^{i+1} = T\{S^i\}$
4, Where $Mask(j) = 1$, update $s^{i+1}(j) := s^0(j)$
5. End for
6. Return $x = s^{iter\_max}$
10. End procedure

---

Firstly, we consider the spectral support is available as a side information. The iterative reconstruction algorithm for this case is as follows [35][7]:

$$x^{k+1}(t) = x^k(t) + \frac{1}{\lambda} f(t) * \{x_s(t) - x_s^k(t)\} \quad (10)$$

where $f(t)$ is the inverse Fourier transform of $F(f)$ defined as:

$$F(f) = \begin{cases} 1 & if\ f \in support\{x\} \\ 0 & else \end{cases}$$

This algorithm has been proposed and proved in [35] for low-pass signals. We restate (Theorem 2 bellow) this algorithm for sparse multi-band signals at the sub-Nyquist Landau rate.

*Theorem 2:* If the signal $x(t)$ is a stationary stochastic process, a sufficient condition for convergence of the iterative relation in (10) to $x(t)$ is:

$$\lambda > 2B \quad (11)$$

where $\lambda$ is the average sampling rate and $2B$ is the total bandwidth of the signal (Landau rate).

The proof of this theorem is given in the Appendix.

If the spectral support of the signal is not known, we need algorithms such as IMAT for its recovery. This algorithm uses adaptive thresholding in each iteration. For a deterministic sparse signal $x(t)$ and random samples $x_s(t)$, the IMAT algorithm has the following iterative relation:

$$x^{k+1}(t) = x^k(t) + \mathcal{F}^{-1} \frac{T^{k+1}}{\lambda} \mathcal{F}\{x_s(t) - x_s^k(t)\} \quad (12)$$

The operator $T^k$ is the threshold operator at the k-th iteration as given below:

$$T^{k+1}\{X(f)\} = \begin{cases} X(f) & if\ |X(f)| \geq Thr^{k+1} \\ 0 & else \end{cases}.$$

$Thr^{k+1}$ is the threshold value which adaptively varies with iteration number.

The following theorem helps us to choose the threshold level properly, to guarantee perfect reconstruction of the algorithm.

*Theorem 3:* Let $x(t)$ be a deterministic signal; if the threshold level is set as:

$$Thr^{k+1} = \alpha \sqrt{\lim_{W \to \infty} \frac{\int_{-W}^{W}[X_s - X_s^k(f)]^2 df}{2W}} \quad (13)$$

and $x^0(t)$ is chosen such that the $support\{x^0(t)\} \subseteq support\{x(t)\}$, then with an arbitrarily large probability, we would have:

$$support\{x^{k+1}(t)\} \subseteq support\{x(t)\}$$

*Proof:* The proof is derived by induction. The basis of induction is true since we assume that $support\{x^0(t)\} \subseteq support\{x(t)\}$. From (12), we have:

$$support\{x^{k+1}(t)\}$$
$$= support\{x^k(t)\}$$
$$\cup support\left\{\frac{T^{k+1}}{\lambda}\{x_s(t) - x_s^k(t)\}\right\}$$

For simplicity, we denote $x(t) - x^k(t)$ by $y^k(t)$. Based on the induction hypothesis, we have:

$$support\{y^k(t)\} \subseteq support\{x(t)\}$$

As previously discussed, in section II, if we choose the threshold level for the signal $y^k(t)$ according to (9):

$$Thr(k+1) \geq \alpha\sqrt{\lambda P_{y^k}} = \alpha\sqrt{\lambda P_{x-x^k}} \quad (14)$$

Then, we would have the following with arbitrarily large probability:

$$support\left\{\frac{T^{k+1}}{\lambda}\{y_s^k(t)\}\right\} \subseteq support\{y^k(t)\}$$

This implies that:

$$support\{x^{k+1}(t)\} \subseteq support\{x(t)\}.$$

According to (3), we have:

$$Y_s^k(f) = \lambda Y^k(f) + N_{Y^k}(f)$$

We can write the following equality as:

$$\lim_{W\to\infty}\frac{\int_{-W}^{W}[X_s-X_s^k(f)]^2 df}{2W} = \lim_{W\to\infty}\frac{\int_{-W}^{W}Y_s^k(f)^2 df}{2W}$$
$$= \lim_{W\to\infty}\frac{\lambda^2\int_{-W}^{W}Y^k(f)^2 df}{2W}$$
$$+ \lim_{W\to\infty}\frac{2\lambda\int_{-W}^{W}Y^k(f)N_{Y^k}(f)df}{2W}$$
$$+ \lim_{W\to\infty}\frac{\int_{-W}^{W}N_{Y^k}(f)^2 df}{2W}$$

The first term is zero since:
$$\int_{-\infty}^{\infty}Y^k(f)^2 df = P_{y^k}.$$

According to Lemma 2, $N_{Y^k}(f)$ is an ergodic process. Hence, we can conclude that:
$$\lim_{W\to\infty}\frac{\int_{-W}^{W}N_{Y^k}(f)^2 df}{2W} = E\{N_{Y^k}(f)^2\} = P_{N_{Y^k}} = \lambda P_{y^k}$$
where the third equality follows from (8).
Likewise, we have:
$$\lim_{W\to\infty}\frac{2\lambda\int_{-W}^{W}Y^k(f)N_{Y^k}(f)df}{2W} = 2\lambda E\{Y^k(f)N_{Y^k}(f)\}$$
$$= 2\lambda Y^k(f)E\{N_{Y^k}(f)\} = 0$$
where the last equality results from Lemma 2. Thus, we have:
$$\lim_{W\to\infty}\frac{\int_{-W}^{W}[X_s-X_s^k(f)]^2 df}{2W} = \lambda P_{x-x^k} \quad (15)$$

As a result, the threshold value is computable as:
$$Thr^{k+1} = \alpha\sqrt{\lim_{W\to\infty}\frac{\int_{-W}^{W}[X_s-X_s^k(f)]^2 df}{2W}}$$
which is equal to the (13) in the Theorem 3 and the proof is thus complete. ∎

Note that the threshold is computable from the sampled signal $x_s$.

*Theorem 4:* The sufficient condition for the iterative relation in (12) to recover any randomly sampled deterministic signal with arbitrarily large probability is:
$$\lambda > \rho 2B \quad (16)$$

where $\lambda$ is the average sampling rate, $2B$ is the total bandwidth of the signal (Landau rate) and $\rho$ (oversampling ratio[1]) is a constant, chosen with respect to the required accuracy.

*Proof:* Define the filter $F^{k+1}$ as follows:
$$F^{k+1}(f) = \begin{cases} 1 & if\ |X_s(f)-X_s^k(f)| > Thr^{k+1} \\ 0 & else \end{cases}$$
and $f^{k+1}(t)$ as its inverse Fourier transform.
The proposed iterative relation in (12) can be rewritten as:
$$x^{k+1}(t) = x^k(t) + \frac{1}{\lambda}f^{k+1}(t)*\{x_s(t)-x_s^k(t)\}$$
$$= x^k(t) + \frac{1}{\lambda}f^{k+1}(t)*\{-y_s^k(t)\}.$$

---
[1] The ratio of the sampling rate divided by the Landau rate is defined as the oversampling ratio.

Thus, $y^{k+1}(t)$ would be derived as:
$$y^{k+1}(t) = y^k(t) - \frac{1}{\lambda}f^{k+1}(t)*\{y_s^k(t)\} \quad (17)$$

We define $h^{k+1}(t)$ as the part of $x-x^k = -y^k$ whose spectrum is restricted to the subset $I^{k+1}$. Also, we suppose $\overline{h^{k+1}}(t)$ as its complement. Therefore, we have:
$$h^{k+1}(t) \stackrel{\text{def}}{=} f^{k+1}(t)*\{x(t)-x^k(t)\} = -f^{k+1}(t)*\{y^k\}$$
$$\overline{h^{k+1}}(t) \stackrel{\text{def}}{=} -y^k(t) + f(t)*\{y^k(t)\}.$$
Hence, $-y^k(t)$ can be written as:
$$-y^k(t) = h^{k+1}(t) + \overline{h^{k+1}}(t) \quad (18)$$

In order to analyze the convergence of the iterative relation, by assuming uniform distribution for random samples, we can have the following relation for the Fourier transform of $y_s^k(t)$ from Theorem 1 and (17):
$$Y_s^k(f) = \lambda Y^k(f) + N_{Y^k}(f)$$
where $N_{Y^k}(f)$ is a white Gaussian noise with the variance:
$$\sigma^2 = P_{N_{Y^k}} = \lambda P_{y^k} \quad (19)$$

Hence, we have:
$$\frac{1}{\lambda}f^{k+1}(t)*y_s^k(t)$$
$$= -h^{k+1}(t) + \frac{1}{\lambda}f^{k+1}(t)*n_{Y^k}(t)$$
$$= -h^{k+1}(t) + z^{k+1}(t) \quad (20)$$
where
$$z^{k+1}(t) = \frac{1}{\lambda}f^{k+1}(t)*n_{Y^k}(t)$$
and

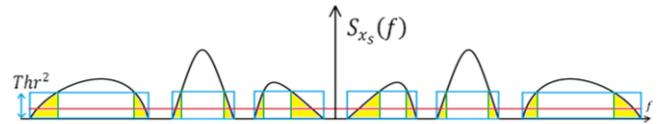

Fig. 2. Power spectrum of the random sampled signal. The hashed area indicats the spectrum region in which the sampled spectrum does not pass the threshold. The horizontal line above the $f$ coordinate indicates the noise level.

$$Z^{k+1}(f) = \frac{1}{\lambda}N_{Y^k}(f)F^{k+1}(f) \quad (21)$$

Thus, (20) can be rewritten as:
$$y^{k+1}(t) = -\overline{h^{k+1}}(t) + \zeta^{k+1}(t)$$
As the frequency support of $\overline{h^{k+1}}$ and $z^{k+1}$ are mutually exclusive, we can have:
$$P_{y^{k+1}} = P_{\overline{h^{k+1}}} + P_{z^{k+1}}$$
According to (19) and (21) and ergodicity of $N_{Y^k}(f)$, we have:

$$P_{z^{k+1}} = \int_{-\infty}^{\infty} \left|\frac{1}{\lambda} N_{Y^k}(f) F^{k+1}(f)\right|^2 df$$
$$= \frac{1}{\lambda^2} P_{N_{Y^k}} \left(\int_{-\infty}^{\infty} F^{k+1}(f) df\right)$$
$$= \frac{1}{\lambda} P_{y^k} \left(\int_{-\infty}^{\infty} F^{k+1}(f) df\right) \quad (22)$$

In order to derive an upper bound for $P_{\overline{h^{k+1}}}$, we first define the filter $\Omega^{k+1}(f)$ as follows:
$$\Omega^{k+1}(f) = \begin{cases} 1 & \text{if } f \in \text{support}\{x(t)\} \text{ and } |X_s(f) - X_s^k(f)| > Thr^{k+1} \\ 0 & \text{else} \end{cases}$$

and its inverse Fourier transform as $\omega^{k+1}(t)$.
If we consider $-y_s^k(t)$ as the random samples of $-y^k(t)$, according to (5), we have:
$$Y_s^k(f) = \lambda Y^k(f) + N_{Y^k}(f)$$

Considering the Fourier transform of (18), we have:
$$Y_s^k(f) = -\lambda H^{k+1}(f) - \lambda \overline{H^{k+1}}(f) + N_{Y^k}(f)$$

Now, define $\overline{h_\omega^{k+1}(t)}$ as follows:
$$\overline{h_\omega^{k+1}(t)} \stackrel{\text{def}}{=} \omega^{k+1}(t) * \{-y_s^k(t)\}$$

Hence, we have:
$$\overline{H_\omega^{k+1}}(f) = \lambda \overline{H^{k+1}}(f) - N_{Y^k}(f) \Omega^{k+1}(f) \quad (23)$$

$\overline{h_\omega^{k+1}(t)}$ consists of the part of $y_s^k(t)$ with the spectrum that lies below $(Thr^{k+1})^2$. For obtaining better intuition, note that $P_{\overline{h_\omega^{k+1}}}$ is the hashed area in Fig. 2.

$$P_{\overline{h_\omega^{k+1}}} = \int_{-\infty}^{\infty} \left|\overline{H_\omega^{k+1}}(f)\right|^2 df \leq Thr^{k+1^2}(2B) \quad (24)$$

The inequality comes from the fact that the area of the hashed regions of a multiband signal in Fig. 2. is less than that of the specified rectangles if we consider $2B$ as the total occupied bandwidth of $x(t)$.

According to Theorem 3 and (23), with arbitrarily large probability, we have:

$$\int_{-\infty}^{\infty} \left|\overline{H_\omega^{k+1}}(f)\right|^2 df$$
$$\geq \lambda^2 \int_{-\infty}^{\infty} \left|\overline{H^{k+1}}(f)\right|^2 df$$
$$- \int_{-\infty}^{\infty} \left|N_{Y^k}(f) \Omega^{k+1}(f)\right|^2 df$$
$$= \lambda^2 P_{\overline{h^{k+1}}} \quad (25)$$
$$- P_{N_{Y^k}} \left(\int_{-\infty}^{\infty} |\Omega^{k+1}(f)| df\right)$$
$$= \lambda^2 P_{\overline{h^{k+1}}}$$
$$- \lambda P_{y^k} \left(\int_{-\infty}^{\infty} \Omega^{k+1}(f) df\right)$$

Considering (24), (25) and non-negativity of $\lambda P_{n^k}(\int_{-\infty}^{\infty} \Omega^{k+1}(f) df)$, we have:

$$\lambda^2 P_{\overline{h^{k+1}}} \leq Thr(k+1)^2(2B) + \lambda P_{y^k} \left(\int_{-\infty}^{\infty} \Omega^{k+1}(f) df\right)$$

Hence

$$P_{\overline{h^{k+1}}} \leq \frac{Thr(k+1)^2(2B) + \lambda P_{y^k}(\int_{-\infty}^{\infty} \Omega^{k+1}(f) df)}{\lambda^2} \quad (26)$$

Using (22) and (26), we conclude that:
$$\frac{P_{n^{k+1}}}{P_{n^k}} = \frac{P_{\overline{h^{k+1}}} + P_{\xi^{k+1}}}{P_{n^k}}$$
$$\leq \frac{(\alpha^2)(2B) + (\int_{-\infty}^{\infty} F^{k+1}(f) df) + (\int_{-\infty}^{\infty} \Omega^{k+1}(f) df)}{\lambda}$$
$$= \frac{(\alpha^2 + 1)(2B)}{\lambda}$$

The last equality is an immediate consequence of the definition of $F^{k+1}(f)$ and $\Omega^{k+1}(f)$. Setting $\lambda > (1 + \alpha^2)(2B)$, we see that $P_{y^k}$ decays with a factor of $\frac{(1+\alpha^2)(2B)}{\lambda}$ which indicates that we have at least a linear convergence of $x^k(t)$ to $x(t)$ in mean squared error sense. Hence, the theorem is proved by defining the oversampling ratio ρ in (16) as:

$$\rho = (1 + \alpha^2) \quad (27)$$
∎

IV. IMPLEMENTATION

In this section, some practical details related to the reconstruction algorithms are discussed and some useful modifications are applied to the proposed method.

*A. The Details of the Reconstruction Algorithm*

To simulate this algorithm on a PC, we use a discrete but highly oversampled signal as an approximation of a real analog signal. We down-sample the input signal with a random sampling mask, and then try to recover the input signal from the samples. Since our algorithm manipulates the signal in time and frequency domain, we can use any orthogonal discrete transform (e.g. DFT, DCT, or DWT). Here, we demonstrate the pseudo-codes for our algorithms given in Theorem 2 and Theorem 4.

---

**Algorithm 2: Iterative Algorithm for Known Support**
- $e, r, m, s$ are vectors with the length $L$ which is a parameter.
- $E(j)$ for any signal $E$ means the index $j$ of the vector $E$ and by $j$ we mean any integer number between 1 and $L$.

**Input:**
$m :=$ Random Sampling Mask
$s^0 :=$ Input Signal Sampled with Mask $M$
$F :=$ Spectral Support of Input Signal
$\lambda :=$ Sampling Rate
**Output:** $r$
1. $R(j) := 0$
3. $S^0 :=$ FFT of $s^0$
4. For $i$ from 0 to number of iterations do
5.      $E := S^i$
6.      Where $j \notin F$, update $E(j) := 0$

7.     $E := E * \lambda^{-1}$
8.     $R := R + E$
9.     $e :=$ Inverse FFT of $E$
10,     Where $m(j) = 0$, update $e(j) := 0$
11.     $E :=$ FFT of $e$
12.     $S^{i+1} := S^i - E$
13. End for
14. $r :=$ Inverse FFT of $R$

**Algorithm 3: A variation of IMAT from Algorithm 1**

- $e, r, m, s$ are vectors with the length $L$ which is a parameter.
- $E(j)$ for any signal $E$ means the index $j$ of the vector $E$ and by $j$ we mean any integer number between 1 and $L$.

**Input:**
$m :=$ Random Sampling Mask
$s^0 :=$ Input Signal Sampled with Mask $m$
$\alpha :=$ Confidence Parameter
$\lambda :=$ Sampling Rate
**Output:** $r$
1. $R(j) := 0$
2. $S^0 :=$ FFT of $s^0$
3. For $i$ from 0 to number of iterations do
4.     $E := S^i$
5.     $thr := \frac{\alpha}{\sqrt{L/2}} \|S^i\|_2$
6.     Where $|E(j)| < thr$, update $E(j) := 0$
7.     $E := E * \lambda^{-1}$
8.     $R := R + E$
9.     $e :=$ Inverse FFT of $E$
10,     Where $m(j) = 0$, update $e(j) := 0$
11.     $E :=$ FFT of $e$
12.     $S^{i+1} := S^i - E$
13. End for
14. $r :=$ Inverse FFT of $R$

As discussed previously in Section II, we set the $\alpha$ value as a constant parameter. Using a trial method, we observe that the value of $\alpha = 2.5$ results in a good performance with a fast convergence rate. Consequently, according to (27), a reliable reconstruction can take place with an oversampling ratio of 7, i.e., a sampling rate of 7 times the Landau rate. We later apply this relation in Subsection IV.B for our parameter settings, and show that the necessary sampling rate may be lower than this sufficient bound.

### B. Modifications and Efficiency Improvement

There may be implementation modifications in practice. For example, it is possible to approximate thresholds with an exponentially damping function with respect to the iteration number to enhance computational complexity. As another example, one can overwrite known samples on corresponding values of $r$, to reach a more accurate result. Another helpful idea is to impose a relaxation parameter $\gamma$ when updating $R$, simply by replacing line 8 (for both algorithms) with $R := R + \gamma * E$. The relaxation parameter is helpful for the faster convergence of the iterative algorithms.

Note that the spectral support of the input signal is determined step by step during iterations of the modified IMAT algorithm (Algorithm 3). Hence, during each iteration, we can use the portion of the spectral support determined at previous iterations to enhance the performance of the new iterations. This simple idea can yield to a new hybrid algorithm by replacing line 6 of Algorithm 3 with the following:
Where $|E(i)| < thr$ and $R(i) = 0$ update $E(i) := 0$

This algorithm is referred to as the Hybrid IMAT in Section V, where we show this slight change highly improves the performance of the reconstruction algorithm.

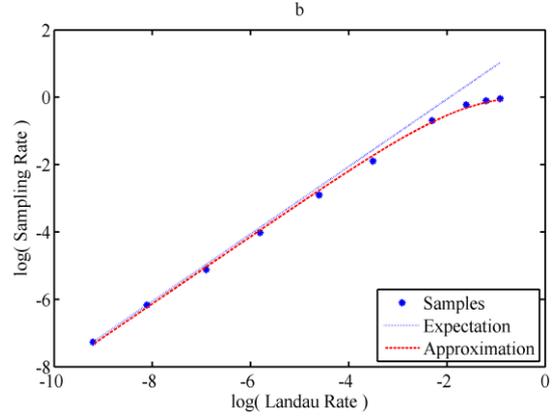

Fig. 3. Perfect (100 dB) reconstruction of Algorithm 3: the required sampling rate versus the Landau rate. The expectation curve is the outcome of the analytical result.

### V. SIMULATION RESULTS

In this section, we study the performance of the proposed methods from different aspects. In Subsection V.A, we evaluate our reconstruction algorithms using general synthetic signals in order to confirm the conditions of perfect reconstruction proved in Section III. The Hybrid IMAT proposed in Section IV is also included in our evaluations. In Subsection V.B, we simulate a noisy multi-band RF signal to evaluate the performance of our hybrid method. We show that the performance of the hybrid method for real signals is even better than the analytical results. We also demonstrate that the proposed method has a denoising capability.

#### A. Simulations of Synthetic Signals

As we proved in Section III, the sufficient condition for perfect reconstruction of our proposed method is an average sampling rate which depends on the Landau Rate. Hence, we use 3 different signals, with identical Landau rates of 30 MHz and the Nyquist rates up to 1 GHz. The applied signals differ in the number and the position of spectral bands. At a sampling rate of 210MHz (i.e., 7 times Landau rate), the IMAT algorithm can reconstruct each of the signals reliably. We consider a 100dB SNR value as a reliable reconstruction.

For consistency of the simulation results, we normalize the sampling and Landau rates by dividing the rates by the Nyquist rate, which is a representation of the sparsity of the signal.

To show experimental sampling rates derived from the simulation results with respect to the Landau rate, we simulated algorithm 3 for synthetic signals at different Landau rates (Fig. 3. ). Our analytical expectation of the required sampling rate (i.e. $R = 7 * L$, from Section IV.A) is also depicted in Fig. 3. This figure confirms the accuracy analytically expected values at low Landau rates. Furthermore, At higher Landau rates, the simulated sampling rates get saturated, and deviates from the analytical results. The reason is that at higher normalized Landau rates the signal becomes dense and the analytical upper bound becomes poor.

Fig. 4. demonstrates the simulation results for Algorithm 2 with a notation similar to Fig. 3. The expectation for this method behaves linearly ($R = L$). In contrast to the IMAT algorithm, the performance of Algorithm 2 is worse than the expectation curve. From this figure we can surmise that when the oversampling ratio is equal to 3 ($R = 3 * L$ Approximation in Fig. 4. ) we get good results.

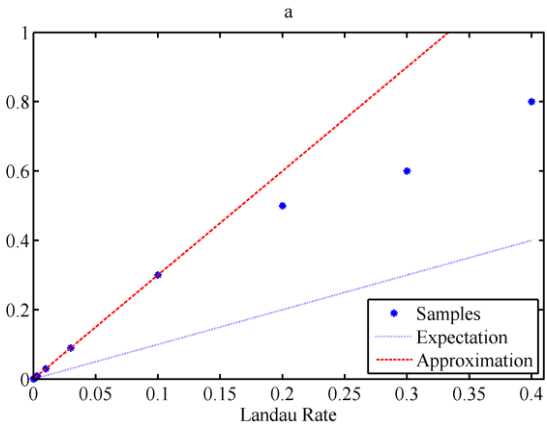

Fig. 4. Perfect (100 dB) reconstruction of Algorithm 2: the required sampling rate versus the Landau rate.

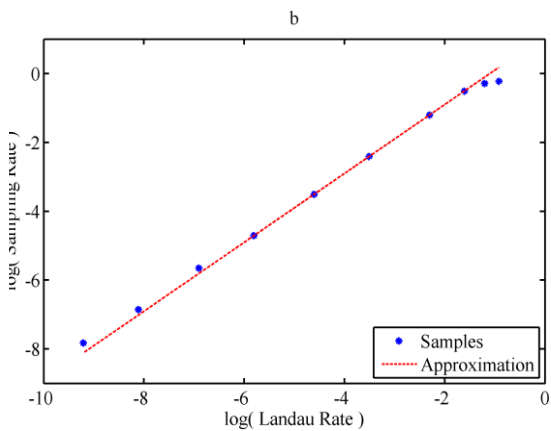

Fig. 5. Hybrid IMAT: required sampling rate for perfect reconstruction (logarithmic view)

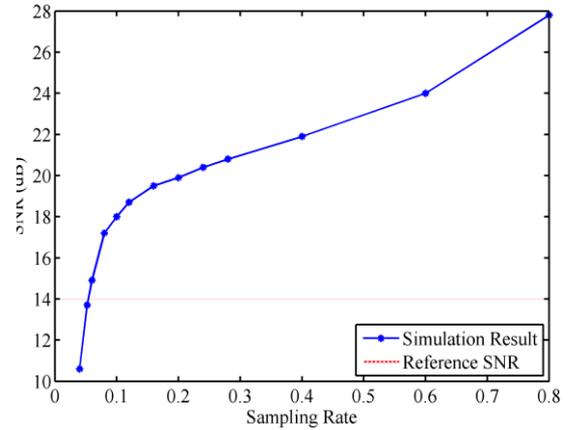

Fig. 6. Simulation results of IMAT on a noisy FM modulated signal

Fig. 5. demonstrates similar evaluation for the Hybrid IMAT. The approximated linear bound for this simulation results can be formulated as $R = 3 * L$, which is as low as Algorithm 2.

*B. Simulation Results for Multi-Band RF Signal*

In this subsection we investigate our method with a real scenario. We use a multi-band RF signal generated by MATLAB. We measure the normalized Landau rate of the generated signal to be approximately 0.04. We add a white Gaussian noise to the signal to simulate a received signal with an SNR value of 14 dB. We randomly sample the received signal with various rates and apply the Hybrid IMAT to reconstruct the sampled signal. Fig. 6. shows the SNR value of the recovered signal for different sampling rates. At $R = 1.3 * L$ (i.e., the sampling rate of 0.05), the Hybrid IMAT algorithm recovers the signal with the SNR equaling that of the received signal. At higher sampling rates, the Hybrid IMAT algorithm shows a denoising effect on the signal, where it recovers up to 18 dB with a sampling rate of 0.1.

VI. CONCLUSION

We have proposed a sub-Nyquist random sampling recovery method for sparse discrete and continuous multi-band signals. Unlike the uniform sampling case, this algorithm does not need an anti-aliasing filter. We provided mathematical proofs for the reliable reconstruction of the iterative recovery algorithms. Furthermore, the simulation results validated the analytical proofs. We showed that our proposed method reliably recovers any signal sampled at least at 3 times the Landau rate. Additionally, this method does not require to know the position of the signal occupied bands in the frequency domain. We also used real RF signals and observed that our system can perform better than the analytical results in practice. We showed that our reconstruction is not only robust against noise, but also has a denoising effect on the input signal.

APPENDIX

Proof of *Lemma2*:

The necessary and sufficient condition for ergodicity of a Gaussian process is the integrablity of absolute value of its autocorrelation function. Since $N(f)$ is a Gaussian process, it will be ergodic if and only if:
$$\int_{-\infty}^{\infty} |R_N(F)| dF < \infty$$
By substituting $R_N(F)$ from (7), it can be rewritten:
$$\int_{-\infty}^{\infty} |R_N(F)| dF = \int_{-\infty}^{\infty} |\lambda X(F) * \overline{X(-F)}| dF$$
$$= \int_{-\infty}^{\infty} \left| \int_{-\infty}^{\infty} \lambda X(f) \overline{X(F-f)} df \right| dF$$
$$\leq \int_{-\infty}^{\infty} \int_{-\infty}^{\infty} |\lambda X(f) \overline{X(f+F)}| df \, dF$$
$$= \int_{-\infty}^{\infty} \int_{-\infty}^{\infty} \lambda |X(f)||X(f+F)| dF \, df$$
$$= \int_{-\infty}^{\infty} \lambda |X(f)| df \int_{-\infty}^{\infty} |X(f+F)| dF$$
$$= \lambda \left( \int_{-\infty}^{\infty} |X(f)| df \right)^2$$

And it follows that the sufficient condition for ergodicity of $N(f)$ is:
$$\int_{-\infty}^{\infty} |X(f)| df < \infty$$
∎

*Proof of Theorem 2:*

We first prove by induction that for each k
$$\text{support}\{x^k(t)\} \subseteq \text{support}\{x(t)\} \quad (28)$$

Assume this is true for k. We conclude this also applies for k+1. From (10), we have:
$$\text{support}\{x^{k+1}(t)\}$$
$$= \text{support}\{x^k(t)\}$$
$$\cup \text{support}\left\{\frac{1}{\lambda} f(t) * \{x_s(t) - x_s^k(t)\}\right\}$$
Thus,
$$\text{support}\{x^{k+1}(t)\}$$
$$= \text{support}\{x^k(t)\}$$
$$\cup [\text{support}\{f\}$$
$$\cap \text{support}\{x_s(t) - x_s^k(t)\}].$$
By considering the definition of filter $F$, it is trivial that:
$$\text{support}\{x^{k+1}(t)\} \subseteq \text{support}\{x(t)\}.$$
To complete the induction, we need to choose $x^0(t)$ such that
$$\text{support}\{x^0(t)\} \subseteq \text{support}\{x(t)\}$$
which is possible for all conditions.

According to (28), $x^k(t)$ can be written as follows:
$$x^k(t) = x(t) + n^k(t)$$
where $\text{support}\{n^k(t)\} \subseteq \text{support}\{x(t)\}$. Hence:
$$f(t) * n^k(t) = n^k(t) \quad (29)$$

With respect to these facts, the iterative method can be written as:
$$x^{k+1}(t) = x^k(t) + \frac{1}{\lambda} f(t) * \{x_s(t) - x_s^k(t)\}$$
$$= x^k(t) + \frac{1}{\lambda} f(t)$$
$$* \{x_s(t) - (x_s(t) + n_s^k(t))\}$$
$$= x^k(t) + \frac{1}{\lambda} f(t) * \{-n_s^k(t)\}$$
Therefore, the iterative method can be written as follows:
$$x^{k+1}(t) = x(t) + n^k(t) - \frac{1}{\lambda} f(t) * \{n_s^k(t)\} \quad (30)$$

where $n_s^k(t)$ contains the random samples of $n^k(t)$.
With respect to (2), the power spectrum of $n_s^k$ can be computed as:
$$S_{n_s^k}(f) = \lambda^2 S_{n^k}(f) + \lambda P_{n^k}$$
According to (29) and considering the definition of filter $F$, the power spectrum of $\frac{1}{\lambda} f(t) * \{n_s^k(t)\}$ will become as follows:
$$S_{\frac{1}{\lambda} f(t) * \{n_s^k(t)\}}(f) = |F(f)|^2 \left\{\frac{1}{\lambda^2} S_{n_s^k}(f)\right\}$$
$$= S_{n^k}(f) + \lambda^{-1} F(f) \{P_{n^k}\}$$
Therefore, we have:
$$\frac{1}{\lambda} f(t) * \{n_s^k(t)\} = n^k(t) - n^{k+1}(t)$$
where the power spectrum of $n^{k+1}(t)$ is equal to $S_{n^{k+1}}(f) = \lambda^{-1} F(f)\{P_{n^k}\}$. By replacing $\frac{1}{\lambda} f(t) * \{n_s^k(t)\}$ in (30), we have:
$$x^{k+1}(t) = x(t) + n^{k+1}(t)$$
We then calculate the power spectrum of $n^{k+1}(t)$ to rewrite the power of sampling noise at iteration k+1 as follows:
$$S_{n^{k+1}}(t) = \lambda^{-1} F(f)\{P_{n^k}\} \Rightarrow P_{n^{k+1}} = \int_{-\infty}^{\infty} S_{n^{k+1}}(f) df$$
$$= \int_{-\infty}^{\infty} \lambda^{-1} F(f)\{P_{n^k}\} df = \frac{2B}{\lambda} P_{n^k}$$
Considering (11), $\frac{2B}{\lambda} < 1$, the power of sampling noise converges to zero exponentially and $x^k(t)$ converges to $x(t)$ and the theorem is thus proved.
∎

ACKNOWLEDGMENT

The authors are grateful to Dr. Mona Ghassemian for useful discussions and for the helpful comments on the first draft of this manuscript. We are also very thankful to Dr. Arash Amini checking the mathematical proofs.